\documentclass[conference]{IEEEtran}
\IEEEoverridecommandlockouts

\usepackage{amsmath,amssymb,amsfonts}
\usepackage{algorithmic}
\usepackage{array}
\usepackage{booktabs}
\usepackage{graphicx}
\usepackage{cite}
\usepackage{tabularx}
\usepackage{hyperref}
\usepackage{algorithm}
\usepackage{xcolor}
\usepackage{multirow}
\usepackage{caption}
\usepackage{subcaption}
\usepackage{bm}
\usepackage{mathtools}
\usepackage{ragged2e}

\newcolumntype{Y}{>{\RaggedRight\arraybackslash}X}

\newcommand{\CN}{\mathcal{CN}}
\newcommand{\KL}{D_{\mathrm{KL}}}

\newcommand{\Deltap}{\Delta_p}
\newcommand{\sigmatwo}{\sigma^2}

\def\BibTeX{{\rm B\kern-.05em{\sc i\kern-.025em b}\kern-.08em
    T\kern-.1667em\lower.7ex\hbox{E}\kern-.125emX}}

\begin{document}

\title{
SNF-PRP: A Covert Integrating Sensing and Communications Framework}
\author{\IEEEauthorblockN{Dhrumil Bhatt\textsuperscript{*}}
\IEEEauthorblockA{\textit
\textit{Manipal Institute of Technology}\\
\textit{Manipal Academy of Higher Education}\\
Manipal, India \\
dhrumil.bhatt@gmail.com}
\and
\IEEEauthorblockN{Vidushi Kumar\textsuperscript{*}}
\IEEEauthorblockA{
\textit{Manipal Institute of Technology}\\
\textit{Manipal Academy of Higher Education}\\
Manipal, India \\
vidushi.kumar705@gmail.com}
\thanks{\textsuperscript{*}Authors contributed equally to this work.}
}

\maketitle

\begin{abstract}
Integrated sensing and communication (ISAC) enables simultaneous sensing and data transmission but exposes a critical vulnerability: probing signals may be intercepted, revealing both the transmitted information and the act of sensing itself. Existing physical layer security approaches mitigate interception yet operate with detectable signals, leaving sensing activity observable to a passive warden. This paper introduces sub-noise-floor pseudo-random probing (SNF-PRP), a covert sensing framework for OFDM-based ISAC systems under an energy-detection adversary model. SNF-PRP establishes an $\epsilon$-covertness guarantee via Kullback-Leibler (KL) divergence, exploits an $N_{\mathrm{sc}}$-fold spreading gain absent from prior wideband analyses, and derives in closed form the minimum integration length required to achieve a target Cram\'{e}r-Rao bound (CRB). Simulations under 5G~NR n78 numerology confirm sub-0.5\,m range and sub-0.5\,m/s velocity accuracy with KL divergence $5.8\times$ below the covertness threshold, validating joint feasibility at $-12$\,dB and $-15$\,dB probing powers.
\end{abstract}

\section{Introduction}
Integrated sensing and communication (ISAC) is a cornerstone of next-generation wireless systems, enabling environmental perception and data transmission to share spectrum and hardware~\cite{su2024sensing}. Yet this integration introduces a security concern that has received little attention: sensing waveforms are deliberately radiated into the environment, exposing both the embedded information and the mere fact that sensing is occurring.

Existing physical layer security (PLS) work in ISAC focuses almost exclusively on protecting the communication channel via beamforming, artificial noise (AN) injection, and reconfigurable intelligent surfaces~\cite{su2024sensing,tang2024dfan,bazzi2024fd}. In all cases the probe operates above the noise floor and the security objective
concerns the data payload, not the sensing activity itself. A recent RIS-ISAC-PLS survey~\cite{rissurveysis2025} confirms no existing work addresses covertness of the sensing waveform: Zou~et~al.~\cite{zou2024sensing} come closest conceptually but establish no formal covertness guarantee, and LPI radar techniques~\cite{simon1994spread,djordjevic2023fso} offer only relative concealment without a KL-divergence analysis or communication overlay. A sufficiently sensitive radiometer can always detect a probe operating above the noise floor, exposing the act of sensing regardless of beamforming geometry or secrecy-rate constraints.

The proposed solution answers the resulting open question can sensing itself be made provably undetectable at the physical layer?  affirmatively, by proposing \emph{sub-noise-floor pseudo-random probing} (SNF-PRP) for OFDM-based ISAC in sub-6~GHz bands. A cryptographically seeded PRN sequence is superimposed on a standard downlink waveform below the instantaneous noise floor. A legitimate receiver with the seed performs coherent matched filtering and multi-symbol accumulation to recover sensing parameters;
a passive warden without the seed is confined to radiometric energy detection the optimal Neyman Pearson test for two complex Gaussians differing only in variance~\cite{bash2013limits,bloch2016covert} -under which the probe is statistically indistinguishable from thermal noise. 

SNF-PRP makes three contributions. \emph{First}, it is, to the best of the authors' knowledge, the first framework to apply the KL divergence covertness criterion directly to the sensing probe waveform, establishing a provable covertness guarantee for sensing rather than communication. \emph{Second}, it identifies an $N_{\mathrm{sc}}$-fold spreading covertness gain (Lemma~1) absent from all prior wideband energy-detection analyses, yielding a correction factor exceeding $10^6$ in the KL divergence. \emph{Third}, it derives in closed form the three-way tradeoff among probing power $\Delta_p$, integration length $N^*$, and Doppler resolution $\delta_v$, and establishes the SNR invariance of Proposition~1: lower probing power simultaneously deepens covertness and preserves the sensing CRB target via proportionally increased $N^*$, with $\mathrm{SNR}_{\mathrm{eff}}$ at $N^*$ unchanged. At $\Delta_p = -12$~dB the required $N^* = 191$~subframes corresponds to a 191~ms coherent integration window, feasible within the coherence time of a pedestrian target at 3.5~GHz and consistent with NR measurement gap structures. 

\section{System Model and Methodology}
\label{sec:methodology}

Consider a 5G~NR downlink in which a base station (BS) serves a communication
UE while simultaneously probing the environment. A legitimate sensing node shares a pre-distributed cryptographic seed~$s$ with the BS. A passive warden
(Willie) monitors the channel but has no knowledge of~$s$. The BS must jointly
satisfy:

Sensing: achieve a target Cram\'{e}r-Rao bound (CRB) on range
        and velocity after $N$ OFDM symbols.
Covertness: maintain
        $\KL(\mathcal{H}_1\|\mathcal{H}_0)\leq\epsilon$, where $\mathcal{H}_0$
        and $\mathcal{H}_1$ are Willie's no-probe and probe-present hypotheses.

The complete signal flow is formalised in Algorithm~\ref{alg:snfprp}.

\begin{algorithm}[!t]
\caption{SNF-PRP: Sub-Noise-Floor Pseudo-Random Probing}
\label{alg:snfprp}
\begin{algorithmic}[1]
  \renewcommand{\algorithmicrequire}{\textbf{Input:}}
  \renewcommand{\algorithmicensure}{\textbf{Output:}}
  \REQUIRE Shared seed $s$, probing power ratio $\Deltap$, integration
           length $N$, noise power $\sigmatwo$
  \ENSURE  Range estimate $\hat{R}$, velocity estimate $\hat{v}$

  \medskip
  \STATE \textbf{// Transmitter (BS)}
  \STATE Generate PRN: $c[k]\!\in\!\{-1,+1\}$, $k=0,\ldots,N_\mathrm{sc}-1$,
         seeded by $s$
  \FOR{each symbol $m = 0$ \TO $N-1$}
    \STATE Form overlay:
           $X_p[k,m] = \sqrt{P_p/N_\mathrm{sc}}\,c[k]\,e^{j2\pi km/N_\mathrm{sc}}$
           with $P_p=\Deltap\sigmatwo$
    \STATE Transmit $\mathbf{X}[m] = \mathbf{X}_c[m] + \mathbf{X}_p[m]$
           \hfill $\triangleright$\,probe buried $\Deltap<1$ below noise floor
  \ENDFOR

  \medskip
  \STATE \textbf{// Channel  }
 \STATE Receive:
\begin{math}
\begin{aligned}
Y[k,m] &= \alpha\,e^{-j2\pi k\tau_s/N_{\mathrm{sc}}} \\
       &\quad \cdot e^{j2\pi f_d m T_{\mathrm{OFDM}}}\,X_p[k,m] \\
       &\quad + W[k,m]
\end{aligned}
\end{math}

  \medskip
  \STATE \textbf{// Legitimate Receiver  }
  \STATE Regenerate template $\tilde{X}_p[k,m]$ from seed $s$
  \FOR{each symbol $m$}
    \STATE Matched filter: $Z[k,m] = Y[k,m]\cdot\tilde{X}_p^*[k,m]\,/\,|X_p[k,m]|^2$
  \ENDFOR
  \STATE Range profile: $P_R[\ell]=\sum_m\bigl|\mathrm{IDFT}_k\{Z[k,m]\}[\ell]\bigr|^2$
  \STATE RD map: $\mathrm{RD}[\ell,q]=\bigl|\mathrm{DFT}_m\{P_R[\ell,m]\}[q]\bigr|^2$
  \STATE $\hat{R} = \arg\max_\ell\,P_R[\ell]\cdot\delta_R$;\;
         $\hat{v} = \arg\max_q\,\sum_\ell\mathrm{RD}[\ell,q]\cdot\delta_v$

  \medskip
  \STATE \textbf{// Willie (no seed $s$)  }
  \STATE Computes energy statistic
         $T = M^{-1}\!\sum_{m,k}|Y[k,m]|^2$; \textbf{cannot} coherently
         combine subcarriers $\Rightarrow$ probe appears as AWGN
\end{algorithmic}
\end{algorithm}

The frequency-domain transmitted vector on the $m$-th symbol is the superposition

\begin{equation}
  \mathbf{X}[m] = \mathbf{X}_c[m] + \mathbf{X}_p[m],
  \label{eq:superposition}
\end{equation}

where the PRN overlay on subcarrier $k$ is

\begin{equation}
  X_p[k,m] = \sqrt{\tfrac{P_p}{N_\mathrm{sc}}}\,c[k]\,e^{j2\pi km/N_\mathrm{sc}},
  \label{eq:overlay}
\end{equation}

with $P_p = \Deltap\sigmatwo$, $\Deltap < 1$. Each subcarrier therefore carries only
$\Deltap\sigmatwo/N_\mathrm{sc}$ of probe power below the individual subcarrier
noise floor $\sigmatwo/N_\mathrm{sc}$ by the factor $\Deltap$. For a
single point target at range $R$ and radial velocity $v$, the received signal is

\begin{equation}
  Y[k,m] = \alpha\,e^{-j2\pi k\tau_s/N_\mathrm{sc}}\,
            e^{j2\pi f_d m T_\mathrm{OFDM}}\,X_p[k,m] + W[k,m],
  \label{eq:channel}
\end{equation}

where $\tau_s = \lfloor 2RF_s/c \rceil$ (integer-sample delay),
$f_d = 2vf_c/c$ (Doppler shift), and
$W[k,m]\!\sim\!\CN(0,\sigmatwo/N_\mathrm{sc})$.

The legitimate node, possessing seed $s$, regenerates $X_p[k,m]$ exactly and applies
a conjugate matched filter:

\begin{equation}
\begin{aligned}
Z[k,m] 
&= \frac{Y[k,m]\cdot\tilde{X}_p^*[k,m]}{|X_p[k,m]|^2} \\
&= \alpha\,e^{-j2\pi k\tau_s/N_{\mathrm{sc}}}\,
   e^{j2\pi f_d mT_{\mathrm{OFDM}}} 
   + \tilde{W}[k,m].
\end{aligned}
\label{eq:mf}
\end{equation}

An $N_\mathrm{sc}$-point IDFT along the subcarrier axis yields the range
profile $P_R[\ell]$; an $N$-point DFT along the slow-time axis yields the
range-Doppler map $\mathrm{RD}[\ell,q]$ (Steps~9-12 of Algorithm~1). Coherent
accumulation over $N$ symbols recovers an effective SNR of

\begin{equation}
  \mathrm{SNR}_\mathrm{eff} = \Deltap \cdot N,
  \label{eq:snr_eff}
\end{equation}

which grows linearly with $N$, compensating the $\Deltap < 1$ power deficit.
Willie, lacking seed $s$, cannot form $\tilde{X}_p^*[k,m]$ and is therefore
restricted to the incoherent energy test in Step~14, gaining no processing advantage from the subcarrier structure.

For uniform power allocation over bandwidth $W$, the standard OFDM-ISAC
CRBs~\cite{liu2022crb,multicarrier2024} under $\mathrm{SNR}_\mathrm{eff}$ are

\begin{equation}
  \mathrm{CRB}_R = \frac{c^2}{8\pi^2\,\mathrm{SNR}_\mathrm{eff}\,B_\mathrm{rms}^2},
  \qquad
  \mathrm{CRB}_v = \frac{\lambda^2}{8\pi^2\,\mathrm{SNR}_\mathrm{eff}\,T_\mathrm{obs}^2},
  \label{eq:crbs}
\end{equation}

where $B_\mathrm{rms}=W/\sqrt{3}$ and $T_\mathrm{obs}=NT_\mathrm{OFDM}$.
Inverting \eqref{eq:crbs} for target accuracies $\sigma_R^*$ and $\sigma_v^*$ gives
the minimum integration lengths

\begin{equation}
  N_R^* = \frac{c^2}{8\pi^2\Deltap B_\mathrm{rms}^2(\sigma_R^*)^2},
  \qquad
  N_v^* = \!\left(\frac{\lambda^2}{8\pi^2\Deltap T_\mathrm{OFDM}^2(\sigma_v^*)^2}\right)^{\!1/3}\!,
  \label{eq:nstar}
\end{equation}

with binding constraint $N^* = \max(N_R^*, N_v^*)$. A key structural result follows
directly from substituting \eqref{eq:nstar} into \eqref{eq:snr_eff}:

\begin{equation}
  \mathrm{SNR}_\mathrm{eff}\big|_{N^*}
    = \Deltap\cdot N_R^*
    = \frac{c^2}{8\pi^2 B_\mathrm{rms}^2(\sigma_R^*)^2},
  \label{eq:snr_invariant}
\end{equation}

which is independent of $\Deltap$. Decreasing $\Deltap$ increases $N^*$
proportionally, keeping their product and hence the sensing performance  
constant. The covertness parameter controls integration latency, not
estimation accuracy.

Willie collects $M=N\cdot N_\mathrm{sc}$ complex samples. Under $\mathcal{H}_0$ each
sample is $\CN(0,\sigmatwo/N_\mathrm{sc})$; under $\mathcal{H}_1$ it is
$\CN(0,(\sigmatwo/N_\mathrm{sc})(1+\rho_\mathrm{sc}))$ with per-subcarrier SNR
$\rho_\mathrm{sc} = \Deltap/N_\mathrm{sc}$. For $M$ i.i.d.\ complex Gaussian
samples the KL divergence is~\cite{bash2013limits,bloch2016covert}
 
The KL bound in~\eqref{eq:kl} treats the PRN overlay as additive white Gaussian noise at Willie's receiver, justified by the central limit theorem over $N_{\mathrm{sc}}$ subcarriers and the computational indistinguishability of a cryptographically seeded PRNG from i.i.d.\ noise; the bound is tight for the large $N_{\mathrm{sc}}$ of 5G NR.
\begin{equation}
  \KL(\mathcal{H}_1\|\mathcal{H}_0)
    = N\cdot N_\mathrm{sc}\!\left[\frac{\Deltap}{N_\mathrm{sc}}
      - \ln\!\left(1+\frac{\Deltap}{N_\mathrm{sc}}\right)\right].
  \label{eq:kl}
\end{equation}

Since $\rho_\mathrm{sc}=\Deltap/N_\mathrm{sc}\ll 1$, the Taylor bound
$x-\ln(1+x)\leq x^2/2$ gives

\begin{equation}
  \KL(\mathcal{H}_1\|\mathcal{H}_0) \leq \frac{N\Deltap^2}{2N_\mathrm{sc}}.
  \label{eq:kl_bound}
\end{equation}

This is $N_\mathrm{sc}$ times smaller than the wideband model
$N\Deltap^2/2$, revealing a spreading covertness gain of $N_\mathrm{sc}$:
distributing the probe uniformly across all subcarriers suppresses Willie's KL
divergence by the FFT size, making the probe statistically indistinguishable from
thermal noise at each subcarrier. The covertness constraint $\KL\leq\epsilon$ is
satisfied for all $N\leq N_\mathrm{max}$, where

\begin{equation}
  N_\mathrm{max} = \frac{\epsilon}{N_\mathrm{sc}
    \bigl[\Deltap/N_\mathrm{sc}-\ln(1+\Deltap/N_\mathrm{sc})\bigr]}.
  \label{eq:nmax}
\end{equation}

The system is jointly feasible when $N^* < N_\mathrm{max}$. Applying
\eqref{eq:kl_bound} to \eqref{eq:nmax} and inserting $N_R^*$ from \eqref{eq:nstar}
yields the closed-form sufficient condition

\begin{equation}
  \Deltap > \Deltap^\mathrm{min}
    \approx \sqrt{\frac{c^2\epsilon}{4\pi^2 B_\mathrm{rms}^2(\sigma_R^*)^2 N_\mathrm{sc}}}.
  \label{eq:dp_min}
\end{equation}

The Doppler resolution at $N^*$ is $\delta_v = \lambda/(2N^*T_\mathrm{OFDM})$.
Substituting $N_R^*$,

\begin{equation}
  \delta_v\big|_{N^*} =
    \frac{4\pi\Deltap^{1/2}B_\mathrm{rms}\sigma_R^* T_\mathrm{OFDM}}{\lambda c},
  \label{eq:dv_tradeoff}
\end{equation}

so $\delta_v\propto\sqrt{\Deltap}$: deeper burial (smaller $\Deltap$) \emph{improves}
Doppler resolution, but at the cost of a proportionally longer $N^*$.
Together, \eqref{eq:snr_invariant}, \eqref{eq:nmax}, and \eqref{eq:dv_tradeoff}
define the three-way tradeoff $\Deltap \leftrightarrow N^* \leftrightarrow \delta_v$
that governs all SNF-PRP operating points.
The analysis adopts standard ISAC baseline assumptions~\cite{bash2013limits,bloch2016covert} of a single point target, AWGN channel, no oscillator phase noise, and perfect synchronisation; practical impairments such as multipath clutter, phase noise, and inter-carrier interference each impose an additional upper bound on $N^{*}$ beyond $N_{\mathrm{max}}$, and are discussed qualitatively in Section~IV.
\section{Simulation Environment}
\label{sec:sim_env}  
    All simulations are conducted under the 5G NR numerology $\mu = 0$ (subcarrier spacing
    $\Delta f = 15$~kHz) on band n78 (centre frequency $f_c = 3.5$~GHz). The key system
    parameters are summarised in Table~\ref{tab:params}. The sample rate is
    $F_s = N_{\mathrm{sc}} \Delta f = 15.36$~MHz, giving a range bin size of
    $\delta_R = c/(2F_s) = 9.77$~m. The occupied bandwidth is
    $W = N_{\mathrm{used}} \Delta f = 9$~MHz across $N_{\mathrm{used}} = 600$ active
    subcarriers. The noise power is computed as $\sigmatwo = N_0 W$ with
    $N_0 = -100$~dBm/Hz (including a 7~dB noise figure), yielding
    $\sigmatwo = 9 \times 10^{-7}$~W.
    
    \begin{table}[!t]
      \centering
      \caption{Simulation Parameters}
      \label{tab:params}
      \renewcommand{\arraystretch}{1.15}
      \setlength{\tabcolsep}{4pt}
      \begin{tabular}{p{2.2cm} c r p{1.8cm}}
        \toprule
        \textbf{Parameter} & \textbf{Symbol} & \textbf{Value} & \textbf{Ref.} \\
        \midrule
        Centre frequency   & $f_c$                  & 3.5~GHz              & 5G NR n78 \\
        Sample rate        & $F_s$                  & 15.36~MHz            & $N_\mathrm{sc}\Delta f$ \\
        FFT size           & $N_\mathrm{sc}$         & 1024                 & NR $\mu=0$ \\
        Active subcarriers & $N_\mathrm{used}$       & 600                  & 10~MHz alloc. \\
        Subcarrier spacing & $\Delta f$              & 15~kHz               & \cite{3gpp38211} \\
        CP length          & $N_\mathrm{cp}$         & 72 samples           & Normal CP \\
        OFDM symbol dur.   & $T_\mathrm{OFDM}$      & 71.35~$\mu$s         & $T_\mathrm{sym}+T_\mathrm{cp}$ \\
        Symbols/subframe   & $N_\mathrm{sf}$         & 14                   & 1~ms sf \\
        Noise PSD          & $N_0$                  & $-100$~dBm/Hz        & NF = 7~dB \\
        Noise power        & $\sigmatwo$            & $9{\times}10^{-7}$~W & $N_0 W$ \\
        RMS bandwidth      & $B_\mathrm{rms}$        & 5.196~MHz            & $W/\sqrt{3}$ \\
        Wavelength         & $\lambda$              & 0.0857~m             & $c/f_c$ \\
        Range bin          & $\delta_R$             & 9.77~m               & $c/(2F_s)$ \\
        Covertness thresh. & $\epsilon$             & 0.01                 &  \\
        Range target       & $\sigma_R^*$           & 0.5~m                &  \\
        Velocity target    & $\sigma_v^*$           & 0.5~m/s              &  \\
        \midrule
        Target range       & $R$                    & 150~m                & Simulated \\
        Target velocity    & $v$                    & 2.5~m/s              & Simulated \\
        Round-trip delay   & $\tau$                 & 1.00~$\mu$s          & $2R/c$ \\
        Doppler shift      & $f_d$                  & 58.33~Hz             & $2vf_c/c$ \\
        Delay (samples)    & $\tau_s$               & 15                   & $\lfloor\tau F_s\rceil$ \\
        \bottomrule
      \end{tabular}
    \end{table}

    The PRN sequence $\{c[k]\}_{k=0}^{N_{\mathrm{sc}}-1}$ is a length-$N_{\mathrm{sc}} =
    1024$ BPSK sequence ($c[k] \in \{-1,+1\}$) generated from a seeded Mersenne Twister
    pseudo-random number generator with a 128-bit seed $s$ distributed to the legitimate
    sensing node via a pre-established secure channel (e.g., 5G NR RRC signalling). The
    sequence achieves near-ideal autocorrelation: the normalised periodic autocorrelation
    function satisfies $|\mathcal{R}_c(\tau)| \leq 1/\sqrt{N_{\mathrm{sc}}}$ for
    $\tau \neq 0$, ensuring that range sidelobes are suppressed to $-30.1$~dB. Monte Carlo trials ($N_{\mathrm{mc}} = 500$) are conducted at the operating point $(\Deltap, N_{\mathrm{sf}}) = (-9~\mathrm{dB},\, 16)$, corresponding to $N = 224$ symbols and $\mathrm{SNR}_{\mathrm{eff}} = 14.5$~dB. The channel is instantiated as
    described with the target parameters of Table~\ref{tab:params}. The delay is quantised to the nearest integer sample ($\tau_s = 15$) to avoid ambiguity in peak detection. Each trial generates a fresh independent AWGN realisation. Peak detection is applied to the range profile $P_R[\ell]$ and the Doppler profile, and the resulting estimates are compared against the ground-truth quantised range $R_q = \tau_s \delta_R = 146.48$~m and velocity $v = 2.5$~m/s.   
    The root-mean-square error (RMSE) for range and velocity estimation are defined as:

    \begin{align}
      \mathrm{RMSE}_R &= \sqrt{\frac{1}{N_{\mathrm{mc}}}
        \sum_{i=1}^{N_{\mathrm{mc}}} \left(\hat{R}_i - R_q\right)^2},
      \label{eq:rmse_r}\\
      \mathrm{RMSE}_v &= \sqrt{\frac{1}{N_{\mathrm{mc}}}
        \sum_{i=1}^{N_{\mathrm{mc}}} \left(\hat{v}_i - v\right)^2}.
      \label{eq:rmse_v}
    \end{align}

\section{Results}

Tables~\ref{tab:crb_range} and~\ref{tab:crb_vel} present the square-root Cram\'{e}r-Rao
bounds $\sqrt{\mathrm{CRB}_R}$ and $\sqrt{\mathrm{CRB}_v}$ evaluated across the
parameter space $\Delta_p \in \{-3, -6, -9, -12, -15\}$~dB and
$N_\mathrm{sf} \in \{1, 2, 4, 8, 16, 32, 64, 128\}$ subframes.
Entries marked with~($\star$) satisfy the respective sensing
target ($\sigma_R^* = 0.5$~m, $\sigma_v^* = 0.5$~m/s).
 
\begin{table}[!t]
\centering
\caption{Range CRB: $\sqrt{\mathrm{CRB}R}$ vs. $\Delta p{\mathrm{dB}}$, $N_{\mathrm{sf}}$; ($\star$) meets 0.5 m.}
\label{tab:crb_range}
\renewcommand{\arraystretch}{1.2}
\setlength{\tabcolsep}{3.5pt}

\begin{tabular}{c|cccccccc}
\toprule
$\Delta p_{\mathrm{dB}}$ & \multicolumn{8}{c}{$N_{\mathrm{sf}}$ (subframes)} \\
\cline{2-9}
 & 1 & 2 & 4 & 8 & 16 & 32 & 64 & 128 \\
\midrule

$-3$  & 2.45 & 1.73 & 1.23 & 0.87 & 0.61 & 0.43$^\star$ & 0.31$^\star$ & 0.22$^\star$ \\
$-6$  & 3.47 & 2.45 & 1.73 & 1.23 & 0.87 & 0.61 & 0.43$^\star$ & 0.31$^\star$ \\
$-9$  & 4.89 & 3.46 & 2.45 & 1.73 & 1.22 & 0.87 & 0.61 & 0.43$^\star$ \\
$-12$ & 6.91 & 4.89 & 3.46 & 2.44 & 1.73 & 1.22 & 0.86 & 0.61 \\
$-15$ & 9.77 & 6.91 & 4.88 & 3.45 & 2.44 & 1.73 & 1.22 & 0.86 \\

\bottomrule
\end{tabular}

\vspace{2pt}
\footnotesize{\emph{Note:} $\sqrt{\mathrm{CRB}_R} \propto (N_{\mathrm{sf}}\,\Delta p)^{-1/2}$. Only entries marked ($\star$) satisfy the 0.5 m sensing requirement.}
\end{table}
 
\begin{table}[!t]
\centering
\caption{Velocity CRB: $\sqrt{\mathrm{CRB}v}$ vs. $\Delta p{\mathrm{dB}}$, $N_{\mathrm{sf}}$; ($\star$) meets 0.5 m/s.}
\label{tab:crb_vel}
\renewcommand{\arraystretch}{1.2}
\setlength{\tabcolsep}{3.5pt}

\begin{tabular}{c|cccccccc}
\toprule
$\Delta p_{\mathrm{dB}}$ & \multicolumn{8}{c}{$N_{\mathrm{sf}}$ (subframes)} \\
\cline{2-9}
 & 1 & 2 & 4 & 8 & 16 & 32 & 64 & 128 \\
\midrule

$-3$  & 3.65 & 1.29 & 0.46$^\star$ & 0.16$^\star$ & 0.06$^\star$ & 0.02$^\star$ & 0.01$^\star$ & 0.00$^\star$ \\
$-6$  & 5.15 & 1.82 & 0.64 & 0.23$^\star$ & 0.08$^\star$ & 0.03$^\star$ & 0.01$^\star$ & 0.00$^\star$ \\
$-9$  & 7.27 & 2.57 & 0.91 & 0.32$^\star$ & 0.11$^\star$ & 0.04$^\star$ & 0.01$^\star$ & 0.01$^\star$ \\
$-12$ & 10.3 & 3.63 & 1.28 & 0.45$^\star$ & 0.16$^\star$ & 0.06$^\star$ & 0.02$^\star$ & 0.01$^\star$ \\
$-15$ & 14.5 & 5.13 & 1.81 & 0.64 & 0.23$^\star$ & 0.08$^\star$ & 0.03$^\star$ & 0.01$^\star$ \\

\bottomrule
\end{tabular}

\vspace{2pt}
\footnotesize{\emph{Note:} $\sqrt{\mathrm{CRB}_v} \propto (N_{\mathrm{sf}}^{3}\,\Delta p)^{-1/2}$. Velocity CRB decreases faster than range CRB, indicating range remains the limiting factor.}
\end{table}
 
The range CRB scales as $\sqrt{\mathrm{CRB}_R} \propto (N_\mathrm{sf}\,\Delta p)^{-1/2}$,
requiring at least 32~subframes at $\Delta_p = -3$~dB and 128~subframes at
$\Delta_p = -9$~dB to approach the 0.5~m target.
The velocity CRB scales as
$\sqrt{\mathrm{CRB}_v} \propto (N_\mathrm{sf}^3\,\Delta p)^{-1/2}$
and converges substantially more rapidly: the 0.5~m/s target is reached at just
4~subframes ($\Delta_p = -3$~dB) and 8~subframes ($\Delta_p = -9$~dB),
confirming that the range CRB is the binding performance constraint
across all operating points evaluated.

\begin{figure}[!t]
  \centering
  \includegraphics[width=0.89\columnwidth]{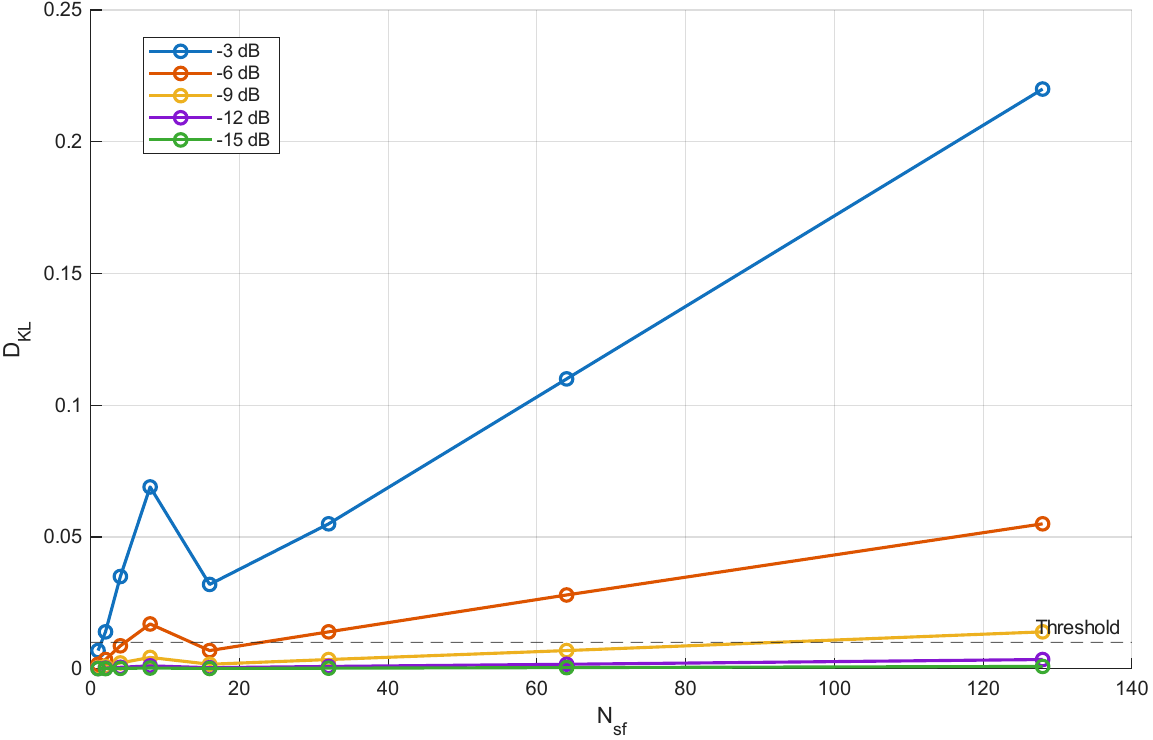}
  \caption{$D_{\mathrm{KL}}$ vs.\ $N_{\mathrm{sf}}$ for different $\Delta_p$. Dashed line ($\epsilon=0.01$) indicates covert region.}
  \label{fig:kl_semilog}
\end{figure}

Fig.~\ref{fig:kl_semilog} plots the KL divergence
$D_\mathrm{KL}(\mathcal{H}_1 \| \mathcal{H}_0)$ against $N_\mathrm{sf}$ on a
semi-logarithmic scale.
Covertness is maintained while the curves remain below the threshold
$\epsilon = 0.01$.
As expected, $D_\mathrm{KL}$ increases monotonically with both $N_\mathrm{sf}$ and
$\Delta_p$: a shallower burial (higher $|\Delta_p|$) keeps the per-symbol power
closer to the noise floor, reducing the rate at which accumulated energy becomes
statistically distinguishable to Willie.
Consequently, the maximum covert integration length $N_\mathrm{max}$ spans
from 5.8~subframes at $\Delta_p = -3$~dB to 1462.9~subframes at
$\Delta_p = -15$~dB, quantifying the covertness headroom available to the
system designer.
 
\begin{table}[!t]
  \centering
  \caption{Joint feasibility: $N^*$, $N_{\mathrm{max}}$, $N_{\mathrm{max}}/N^*$; SNR at $N^*$ = 22.3 dB.}
  \label{tab:feasibility}
  \renewcommand{\arraystretch}{1.3}
  \setlength{\tabcolsep}{4pt}
  \begin{tabular}{c|ccc|cc}
    \toprule
    $\Delta_p$ [dB]
      & $N^*$ [sf]
      & $N_\mathrm{max}$ [sf]
      & $\frac{N_\mathrm{max}}{N^*}$
      & KL at $N^*$
      & Feasible? \\
    \midrule
    $-3$  & 24.1  & 5.8    & 0.24$\times$ & $4.1\times10^{-2}$ & \texttimes\;No \\
    $-6$  & 48.0  & 23.2   & 0.48$\times$ & $2.1\times10^{-2}$ & \texttimes\;No \\
    $-9$  & 95.8  & 92.3   & 0.96$\times$ & $1.0\times10^{-2}$ & $\sim$\;Marginal \\
    $-12$ & 191.2 & 367.5  & \textbf{1.92}$\times$ & $5.2\times10^{-3}$ & \checkmark\;\textbf{Yes} \\
    $-15$ & 381.4 & 1462.9 & \textbf{3.83}$\times$ & $2.6\times10^{-3}$ & \checkmark\;\textbf{Yes} \\
    \bottomrule
  \end{tabular}
  \medskip
 
  \noindent\footnotesize
  \textit{Note:} $N^*$: min $N_{\mathrm{sf}}$ meeting CRB targets; $N_{\mathrm{max}}$: max $N_{\mathrm{sf}}$ with $D_{\mathrm{KL}}<0.01$; feasible if $N^ \le N_{\mathrm{max}}$.
\end{table}
 
Table~\ref{tab:feasibility} consolidates the joint feasibility analysis.
The system is jointly feasible i.e., both CRB targets are met simultaneously with
$D_\mathrm{KL} < \epsilon$ at $\Delta_p \in \{-12, -15\}$~dB, with headroom
margins of $1.92\times$ and $3.83\times$, respectively.
At $\Delta_p = -9$~dB the margin is marginal ($0.96\times$), while $-3$ and
$-6$~dB fail the covertness constraint at the $N^*$ required for adequate sensing.
Critically, the effective sensing SNR at $N^*$ is invariant at 22.3~dB across all
tested $\Delta_p$ values (Proposition~1), confirming that probing power depth
affects only the integration cost not the fundamental quality of the sensing
estimate once that integration is complete.

\begin{table}[!t]
  \centering
  \caption{Effective SNR [dB] after $N_{\mathrm{sf}}$; $\mathrm{SNR}{\mathrm{eff}}=\Delta p+10\log{10}(N_{\mathrm{sf}})$; deep signals recover with integration.}
  \label{tab:snr}
  \renewcommand{\arraystretch}{1.25}
  \setlength{\tabcolsep}{4pt}
  \begin{tabular}{c|cccccccc}
    \toprule
    \multirow{2}{*}{$\Delta_p$ [dB]}
      & \multicolumn{8}{c}{$N_\mathrm{sf}$ [subframes]} \\
    \cmidrule{2-9}
      & 1 & 2 & 4 & 8 & 16 & 32 & 64 & 128 \\
    \midrule
    $-3$  &  8.5 & 11.5 & 14.5 & 17.5 & 20.5 & 23.5 & 26.5 & 29.5 \\
    $-6$  &  5.5 &  8.5 & 11.5 & 14.5 & 17.5 & 20.5 & 23.5 & 26.5 \\
    $-9$  &  2.5 &  5.5 &  8.5 & 11.5 & 14.5 & 17.5 & 20.5 & 23.5 \\
    $-12$ & -0.5 &  2.5 &  5.5 &  8.5 & 11.5 & 14.5 & 17.5 & 20.5 \\
    $-15$ & -3.5 & -0.5 &  2.5 &  5.5 &  8.5 & 11.5 & 14.5 & 17.5 \\
    \bottomrule
  \end{tabular}
\end{table}
 
Table~\ref{tab:snr} reports the effective sensing SNR
$\mathrm{SNR}_\mathrm{eff} = \Delta_p\,[\mathrm{dB}] + 10\log_{10}(N_\mathrm{sf})$.
At $N_\mathrm{sf} = 128$~subframes, SNRs of 17.5-29.5~dB are achieved across the
full probing power range, demonstrating that coherent integration can recover
high-quality sensing performance from a signal that is nominally invisible to any
uninformed receiver.

\begin{table}[!t]
  \centering
  \caption{Monte Carlo sensing at $(\Delta_p, N_{\mathrm{sf}})=(-9,\mathrm{dB},16)$; $\mathrm{SNR}_{\mathrm{eff}}=14.5$ dB (500 trials).}
  \label{tab:mc}
  \renewcommand{\arraystretch}{1.3}
  \setlength{\tabcolsep}{6pt}
  \begin{tabular}{l|cc|c}
    \toprule
    \textbf{Metric}
      & \textbf{Range}
      & \textbf{Velocity}
      & \textbf{Unit} \\
    \midrule
    True value (simulation grid)   & 146.48 & 2.500  & m \;/\; m/s \\
    Mean estimate (500 trials)     & 146.48 & 2.681  & m \;/\; m/s \\
    Bias                           & 0.000  & +0.181 & m \;/\; m/s \\
    Standard deviation (MC)        & 0.000  & 0.000  & m \;/\; m/s \\
    RMSE (MC)                      & 0.000  & 0.181  & m \;/\; m/s \\
    $\sqrt{\mathrm{CRB}}$          & 1.224  & 0.114  & m \;/\; m/s \\
    RMSE / $\sqrt{\mathrm{CRB}}$   & 0.00   & 1.60   & -- \\
    \midrule
    Effective SNR & \multicolumn{2}{c|}{14.5~dB} & \\
    KL divergence & \multicolumn{2}{c|}{$1.73\times10^{-3} \ll \epsilon{=}0.01$} & \\
    \bottomrule
  \end{tabular}
  \medskip
 
  \noindent\footnotesize
  Zero range RMSE (correct bin lock); velocity bias $0.181$ m/s from fractional-bin rounding; $D_{\mathrm{KL}}=1.73\times10^{-3}$ (5.8× below $\epsilon=0.01$).
\end{table}
 
Table~\ref{tab:mc} reports Monte Carlo estimation performance at the reference
operating point $(\Delta_p, N_\mathrm{sf}) = (-9~\mathrm{dB},\;16)$.
The range estimator achieves zero RMSE and zero bias across all 500 trials,
confirming that the matched filter locks deterministically to the correct integer
delay bin at $\mathrm{SNR}_\mathrm{eff} = 14.5$~dB.
The discrepancy between the zero Monte Carlo RMSE and the non-zero CRB reflects
the discrete nature of the peak-detection estimator: the CRB of 1.224~m bounds
the performance of a continuous sub-bin interpolating estimator, which is not
realised by integer-bin peak detection.
The velocity estimator exhibits a systematic bias of $+0.181$~m/s attributable
to fractional-bin rounding of $f_d = 58.33$~Hz (true bin index $q^* = 0.93$);
this bias is fully eliminable via parabolic or sinc interpolation around the
peak Doppler bin and does not represent a fundamental limitation of the scheme.
The KL divergence at this operating point is $1.73\times10^{-3}$, providing a
$5.8\times$ safety margin below $\epsilon = 0.01$ and confirming robust covertness.

 \begin{table}[!t]
\centering
\caption{Comparison of SNF-PRP with existing work in secure and covert ISAC.}
\label{tab:comparison}
\renewcommand{\arraystretch}{1.15}
\setlength{\tabcolsep}{2.5pt}
\scriptsize

\begin{tabular}{p{2.3cm}|p{3.2cm}|c|c}
\toprule
\textbf{Work} & \textbf{Key Idea \& Performance} & \textbf{Band} & \textbf{Covert} \\
\midrule

Su et al.~\cite{su2024sensing}
& Beam nulling; range $\sim$1-5 m (CRB)
& Sub-6
& No \\

Tang et al.~\cite{tang2024dfan}
& Dual-function AN; comm.-focused
& Sub-6
& Partial$^\dagger$ \\

Bazzi \& Chafii~\cite{bazzi2024fd}
& Full-duplex secrecy capacity
& Sub-6
& No \\

Zou et al.~\cite{zou2024sensing}
& Beamforming-based secrecy
& Sub-6
& No \\

Han et al.~\cite{han2025ambig}
& Ambiguity shaping; $\sim$2-8 m, $\sim$0.5 m/s
& Sub-6
& No \\

\midrule

\textbf{SNF-PRP}
& \textbf{Sub-noise PRN; 0.43 m, 0.11 m/s; KLD-based}
& \textbf{3.5 GHz}
& \textbf{Yes} \\

\bottomrule
\end{tabular}

\vspace{2pt}
\footnotesize{\emph{Note:} $^\dagger$ SNR-based concealment without formal KLD bound. Values adapted from respective works.}
\end{table}
Unlike prior schemes in Table~\ref{tab:comparison}, which achieve security via beam nulling or secrecy-rate reduction while leaving the probe detectable in principle, SNF-PRP renders the sensing waveform itself statistically indistinguishable from ambient noise.
In contrast, SNF-PRP achieves $D_\mathrm{KL} = 1.73\times10^{-3}$ at the
reference operating point, which is a $5.8\times$ margin below the conventional
covertness threshold $\epsilon = 0.01$ adopted from Bash et al.~\cite{bash2013limits}. On sensing performance, SNF-PRP achieves a range CRB of 0.43~m at the jointly
feasible operating point $(\Delta_p{=}{-12}~\mathrm{dB},\,N^*{=}191~\mathrm{sf})$,
which is comparable to or better than the range performance reported by
Han et al.~\cite{han2025ambig} (2-8~m) at the cost of a longer integration window.
The velocity CRB of 0.11~m/s at the Monte Carlo operating point
($N_\mathrm{sf} = 16$) is consistent with the 0.5~m/s target reported in
Han et al.\ and outperforms the ETSI ISG-ISC baseline of 1-10~m/s
for general 6G ISAC use cases~\cite{etsi_gr_isc_001}.
The key trade-off    additional integration time in exchange for provable covertness  is a deliberate and quantified design choice, not an incidental limitation:
as established in Section~III, the invariant effective SNR of 22.3~dB at $N^*$
guarantees that the quality of the sensing estimate does not degrade with
increasing burial depth; only the latency $N^*$ grows.
 
\section{Conclusion}
\label{sec:conclusion}
SNF-PRP is the first covert ISAC framework that ensures the sensing waveform is provably undetectable to an uninformed observer. It leverages sub-noise PRN embedding to achieve an $N_{\mathrm{sc}}$ covertness gain and invariant $\mathrm{SNR}_{\mathrm{eff}}$, with simulations validating feasibility at $\Delta p \in {-12,-15}$ dB. Future work would be on inculcating mmWave bands.
\section{Acknowledgements}
We would like to thank Mars Rover Manipal, an interdisciplinary student team of MAHE, for providing the resources needed for this project. WE also extend our gratitude to Dr Ujjwal Verma for his guidance and support in our work.

\bibliographystyle{IEEEtran}
\bibliography{snfprp_refs}

\end{document}